\documentstyle[aps,pre,multicol,epsf]{revtex}
\begin{document}
\draft
\date{April 9, 2000}
\title{Cluster Persistence: a Discriminating Probe of Soap Froth Dynamics}
\author{W. Y. Tam (1), A. D. Rutenberg\thanks{Permanent address:
Department of Physics, Dalhousie University, Halifax NS, Canada B3H 3J5}
(2), B. P. Vollmayr-Lee (3), and K.Y. Szeto (1)}
\address{
(1) Department of Physics, The Hong Kong University of Science and
Technology, Clear Water Bay, Kowloon, Hong Kong\\
(2) Physics Department, 
McGill University, Montr\'{e}al QC, Canada H3A 2T8\\
(3) Department of Physics, Bucknell University, Lewisburg, PA 17837, USA
}
\maketitle
\begin{abstract}
The persistent decay of bubble clusters in coarsening two-dimensional 
soap froths is
measured experimentally as a function of cluster volume fraction.
Dramatically stronger decay is observed in comparison to soap froth models
and to measurements and calculations of persistence in other systems. The fraction
of individual bubbles that contain any persistent area also decays, implying
significant bubble motion and suggesting that T1 processes play an important
role in froth persistence.
\end{abstract}

\begin{multicols}{2}

Cellular structures are common in nature \cite{Weaire84,Glazier92,Stavans93}, 
and are studied in soap froths, polycrystallites, Potts models, biological cell clusters,
and magnetic bubble arrays. They are composed of cells with
varying sizes but identical composition, packed tightly against neighbors
so that cell boundaries meet at vertices.  In most cases, this structure is 
{\em dynamic}, i.e., part of an ongoing nonequilibrium process,
and the challenge
is to understand the role of dynamics in determining the structure.  Individual
cells continually evolve so as to reduce the total interfacial area in 
the system, which results in {\em coarsening}, or growth of the
characteristic bubble size.
``Topological'' changes of cell adjacencies occur when neighbors swap 
($T1$ process) and when cells are destroyed ($T2$ process) \cite{Levitan96}.
Topological transitions are effectively instantaneous in soap froths, while
bubble areas grow smoothly.  This is in contrast to polycrystallites
or dynamical Potts models where topological processes occur at the
same slow time scales as the evolution of bubble areas. 
Rapid topological transitions in soap froths allow us, in principle, to 
isolate the effects of these processes. 

Soap froths are ideal systems to study cellular pattern evolution,
 since they coarsen at laboratory time scales and their time-dependent 
structure can be directly imaged in two-dimensions ($2d$) with 
video microscopy \cite{Glazier87}.  In this paper, we study isotropic froths in $2d$.
We limit ourselves to the scaling regime \cite{Stavans90} at late
times, when the absence of non-universal transients facilitates both the analysis 
and the comparison with model systems. 

A relatively new and general  probe of non-equilibrium dynamics is the local
power-law decay of persistence towards zero \cite{persistence},
$P(t) \sim t^{-\theta}$, which in 
froths measures the fraction of the system that has always 
remained within the same bubble from time $t_0$ through to a later time 
$t \gg t_0$.  A simple theoretical argument provides a lower bound $\theta \geq 1$:
the area of remaining bubbles
in a coarsening froth increase linearly in time, implying that the number
of bubbles $N \sim 1/t$ \cite{Glazier92,Stavans93}. 
Since persistent volume cannot increase within a given
bubble, then $P$ is bounded above by $N \sim 1/t$ asymptotically. 
Numerical studies of $2d$ cellular systems have found $\theta \leq 1$
\cite{Derrida96,Levitan97,Hennecke97}, and
are taken in conjunction with this theoretical bound to indicate $\theta=1$ asymptotically
\cite{Hennecke97}.
However, experimental studies of $2d$ soap froths found
$\theta>1$ \cite{Tam97}, {\em in clear disagreement}, though
consistent with the theoretical bound of $\theta \geq 1$.
Evidently, persistence probes interesting physics in soap froths.

Most studies of froths use measures specific to the cellular topology, 
such as the distribution  of the number vertices per cell.  
While such measures are important (see, e.g.,
\cite{Aste96}), it is also useful to develop probes of structure that are 
{\em not unique to cellular systems}.   Results can then
be compared to corresponding systems that have no cellular structure
and hence no topological processes, so that the distinct effects of
topological dynamics can be revealed.  
For example, it is interesting to compare froths with 
non-conserved scalar coarsening since they
both exhibit curvature-driven interface motion while the
latter has no topological processes.  However, the choices for such probes
are limited, since froth correlations generally
reflect the cellular structure (see, e.g., \cite{Szeto98})
independently of the dynamics. 

Two of us have proposed a more general definition of persistence that makes use
of the soap froth structure to create a
``non-topological'' measure that can be compared to other systems 
\cite{Lee97,endnote1}.  One first randomly selects a
fraction $\phi$ of all of the bubbles at an initial time $t_0$ and colors
them ``green.''  One then asks what fraction of the system has {\em always}
been green between $t_0$ and $t$.   Essentially, we first
make a virtual phase and then study its persistence properties. This
allows us to fingerprint the dynamics at different length scales,
as probed by the size of clusters of the virtual phase.  The limit $\phi=0^+$
consists of isolated bubbles whose persistence decays with the encroachment
of any neighbor, which corresponds to the persistent decay previously
studied in froths. Phases with $\phi>0$ have larger clusters, and for 
sufficiently large $\phi$ these clusters percolate through the system.
Since $\phi$ is also the volume fraction, this ``cluster
persistence'' can be directly compared to persistence in a two-phase system
at the same volume fraction but lacking an underlying cellular structure. 

Our primary results reported in this paper include (i) {\em experimental} 
measurements of cluster persistence for {\em all} volume fractions 
$0 < \phi \leq 1$ in a coarsening $2d$ soap froth; for all 
$\phi$ we find a power law decay with exponent $\theta(\phi)$;
(ii) these exponents $\theta(\phi)$ are greater at corresponding $\phi$
than any other studied system with curvature-driven growth, cellular or non-cellular; 
(iii) the persistence exponent is almost linear with $\phi$, for which we 
provide a mean-field argument; 
and (iv) as $\phi\to 1$, $\theta(\phi)$ 
vanishes linearly with a considerably larger slope than predicted \cite{Lee97} 
for a mean-field droplet model without cellular structure.   We 
interpret these results, particularly point (ii), as evidence of the
strong role of T1 processes in froths.

The experimental setup has been reported earlier \cite{Tam97,Tam98}.
We start our experiment by pumping approximately $20\ 000$ soap bubbles
into a chamber formed by two rigid parallel plates
($26.7\ \times\ 36.8$ cm$^2$ with a $0.16$ cm gap), making an initially
random froth that proceeds to coarsen.  The froth was well drained before 
starting the experiment. Images of the froth were captured using
a high resolution CCD camera ($1037 \times 1344$ pixels) every $10$ min,
sufficient to follow the evolution of individual bubbles.
[The entire chamber is slightly larger than the imaged region.  We
also excluded strips of bubbles near the edge of our image region to test boundary
effects: none were found.] The scaling regime, as identified by a stationary
distribution of bubble topologies, set in from $6$ h when there
were approximately $6000$ bubbles, and we continued taking data until $160$ h
when $200$ bubbles remained.  Our persistence data was taken from $t_0=12$ h, 
with $2000$ bubbles. Quantitatively consistent results, though 
noisier, are obtained using $t_0=26$ h.

\begin{figure}
\begin{center}
\hbox{
\epsfxsize=2.8truein
\epsfbox{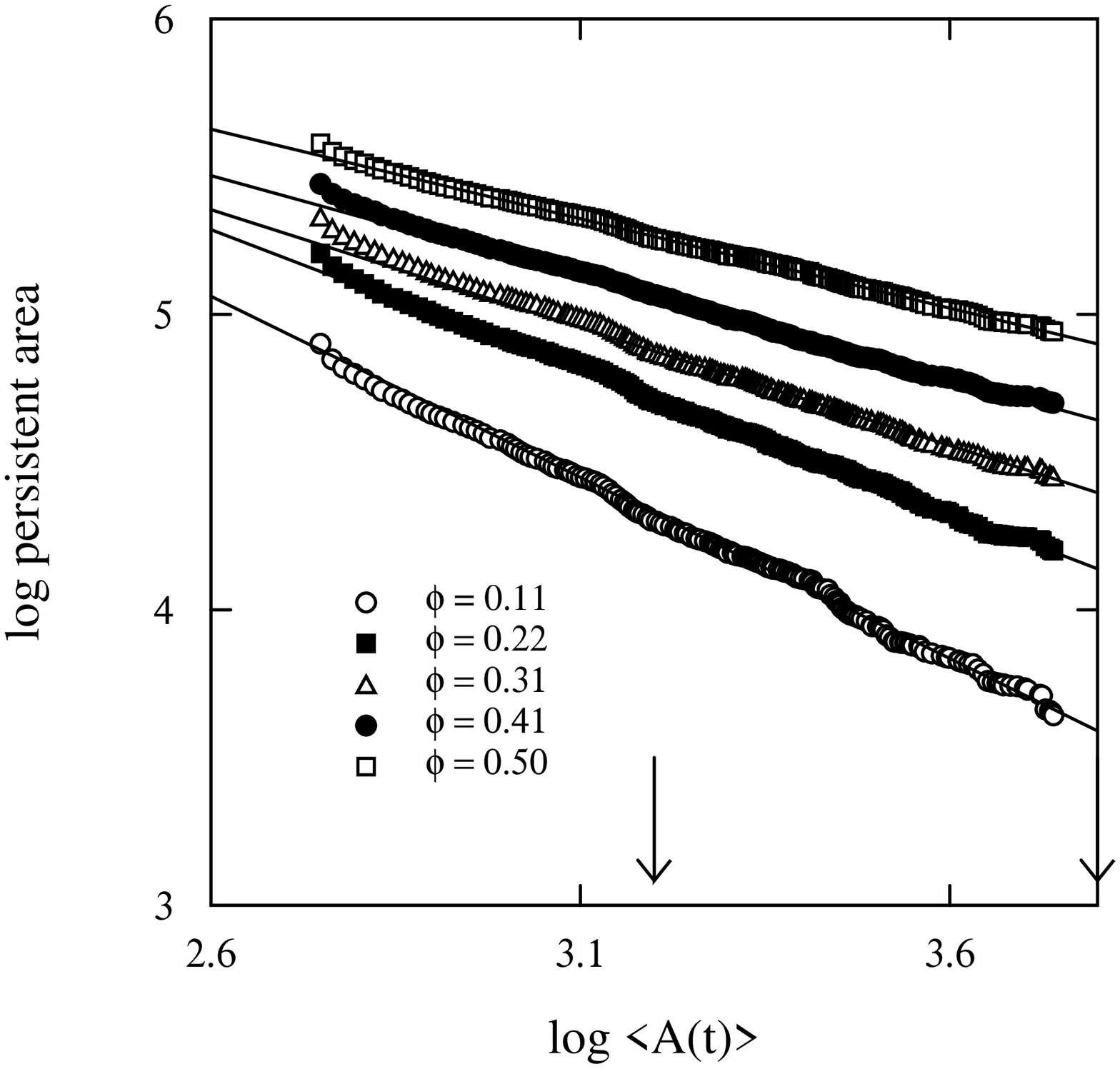}
\qquad
\epsfxsize=2.8truein
\epsfbox{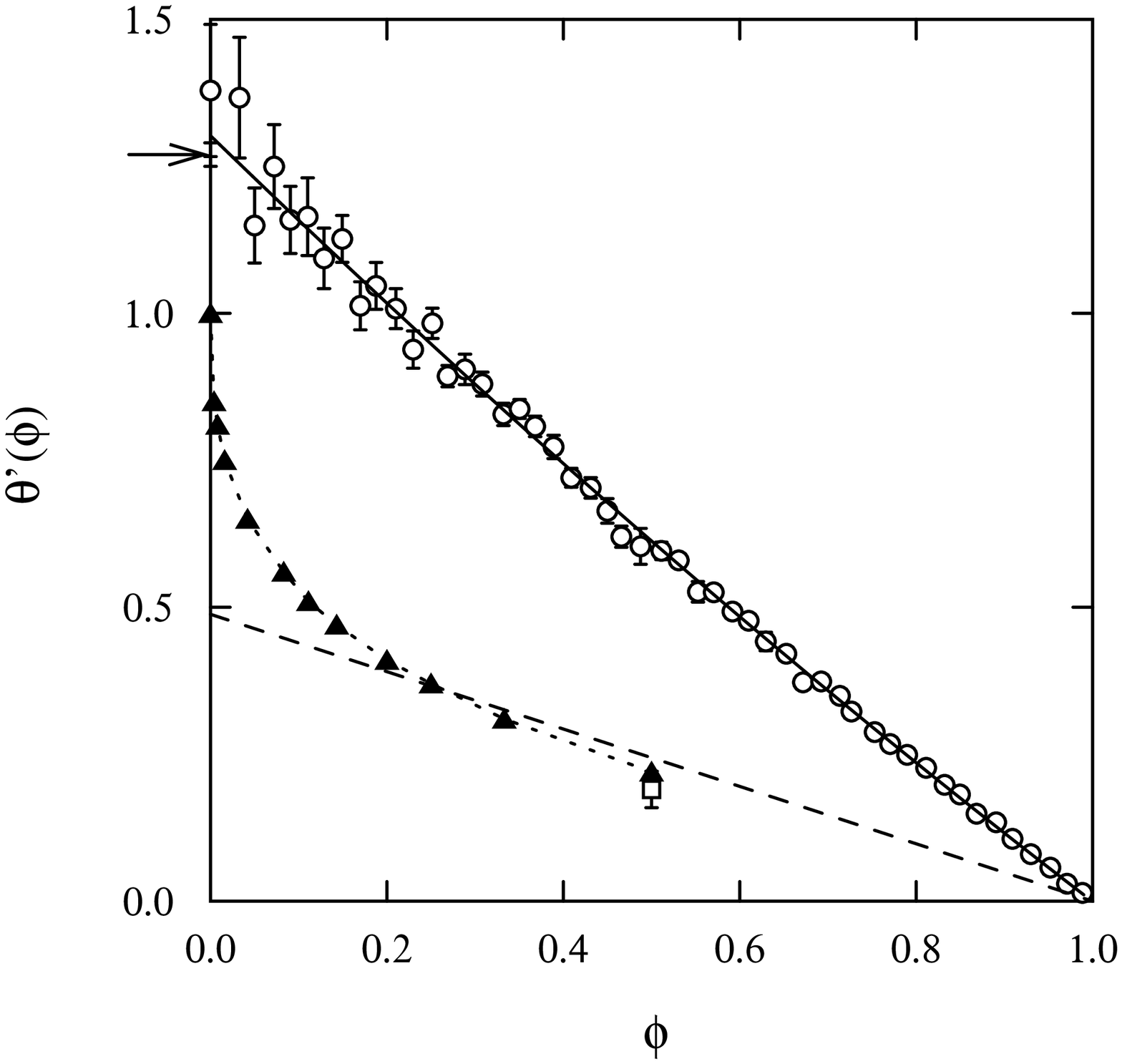}
}
\end{center}
\caption{Left: log-log [base $10$] 
plot of cluster persistence decay for various volume fractions
$\phi$.  The arrows indicate the fitting region used to extract all 
persistence 
exponents. Right: persistence decay exponent as a function of volume fraction
for our soap-froth data (open circles), Potts models with $\phi=1/p$
(triangles),  twisted nematics (open square), and droplet calculation 
(dashed line).  The solid curve is a quadratic fit [see text], while
the dotted curve is a guide to the eye. The arrow indicates the $\phi=0$ value
expected from bubble motion, as probed by 
the decaying number of bubbles with persistent cores [see text and fig.~2]. }
\label{FIG:exppersist}
\end{figure}

We analyze the CCD images to identify each pixel with a given bubble (see
\cite{Tam98}).  Boundary pixels, on bubble boundaries, 
are randomly assigned to adjacent bubbles.
For a given subset $\cal S$ of bubbles, we represent the pixels in $\cal S$ 
by a function $p_{\cal S}(j,t)=1$, where $j \in [1,1393728]$ counts the $M$ 
pixels, and for pixels outside of that subset of bubbles, $p_{\cal S}(j,t)=0$.
The cluster persistence of the set $\cal S$ is then formally
\begin{equation}
        P_{\cal S}(t)
                = \frac{1}{M}\sum_{j=1}^M \prod_{t'=t_0}^t p_{\cal S}(j,t'),
\end{equation}
where $0 \leq P_{\cal S} \leq 1$.
In other words, a pixel counts towards the persistence if it has always been
in any of the bubbles belonging to $\cal S$.  With this definition, the
traditional measure of persistence corresponds to the case where $\cal S$ consists
of a single bubble.  More generally we can take arbitrary subsets of all of
the $N(t_0)$ bubbles, where there are $2^{N(t_0)}$ possible subsets.  We sample
subsets by randomly choosing $m = \phi N(t_0)$ bubbles at a time.  We fit the persistent
decay for individual subsets, and then average the resulting exponents ---
binning them with respect to $\phi({\cal S})$.
For each subset, we use the time-average of $\phi({\cal S})$. 
The resulting persistence decay is characterized by a power-law
\begin{equation}
\label{EQN:theta}
        P_\phi(t) \sim t^{-\theta(\phi)}
                        \sim \langle A \rangle ^{-\theta'(\phi)}.
\end{equation}
We expect $\theta'=\theta$, since the average bubble area in the 
froth, $\langle A \rangle$, is asymptotically proportional to $t$
\cite{Stavans93}, but we measure $\theta'$ to eliminate the dependence
on the absolute time origin and to minimize transient effects
\cite{Hennecke97,Tam98}.  
The decay is well fit to a power-law, as illustrated
for five subsets in the left of fig.~\ref{FIG:exppersist}, and we
plot all of the resulting exponents 
in the right of fig.~\ref{FIG:exppersist} as a function of volume fraction, $\phi$.
The errors shown are from the statistical variation between subsets at a given
volume fraction, and can be reduced with more sampling. We selected about $1000$ 
multiple-bubble subsets out of the $2^{2000}$ possibilities, in addition to the 
$2000$ single-bubble subsets.

In the right of fig.~\ref{FIG:exppersist}
we plot our data for $\theta(\phi)$ as open circles, and
include a quadratic fit as a solid curve (described below).  
For comparison, we include data from 
previous studies of corresponding systems with curvature-driven growth and volume fraction 
$\phi$.  These include the persistence of one phase of the p-state Potts 
model, with $\phi=1/p$, simulated \cite{Derrida96,Levitan97} and plotted as triangles
for $0< \phi \leq 1/2$. 
Potts systems, particularly when $p \rightarrow \infty$,
have been proposed as soap froth analogues \cite{Glazier90}. 
The Potts data fits relatively well on a continuous curve 
(dotted line), interpolating smoothly between $p=2$ with a two-phase non-cellular
structure, $3 < p<\infty$ with a cellular structure and some bubble
coalescence, and $p=\infty$ with a cellular structure and 
no bubble coalescence.  Also shown (open square) is a
previous persistence experiment on twisted nematic liquid crystals
\cite{Yurke97}, an Ising model analog with $\phi=1/2$.  Finally, an analytic 
result for a mean-field droplet model, valid for $\phi \approx 1$, 
describing non-conserved order parameter dynamics with a time-dependent
applied field is plotted as a dashed line
\cite{Lee97}.  We note that remarkable agreement is seen
between the Potts data, the experimental work on twisted nematics,
and the extrapolation of the $O(1-\phi)$  exact
result for coarsening droplets.
Neither the nematic nor the droplet system have cellular structures, and the
relatively good agreement between them and the Potts data indicates
that persistence is not greatly affected by the cellular structure 
{\em per se}.
Nevertheless, we find froth decay exponents that are dramatically larger
than all these corresponding systems!

\end{multicols}

\begin{multicols}{2}

Qualitatively, the froth persistence data is strikingly linear. 
The droplet model studied in Ref.~\cite{Lee97} used a 
mean-field treatment, asymptotically correct as $\phi\to 1$, 
to obtain the linear variation $\theta(\phi) \simeq 0.48797 (1-\phi)$.  This motivates a 
similar mean-field argument for froths.   In essence,
if the local environment around a persistent region is entirely random,
then the rate of  persistent volume lost from a subset of bubbles $\cal S$  of 
volume fraction $\phi$ is given simply by the loss of persistent volume from 
individual bubbles times the chance, $1-\phi$,
that it is being lost to a bubble outside of $\cal S$.  That is,
$\dot{P}_\phi/P_\phi = (1-\phi) \dot{P}_{0}/P_0$
where $P_\phi$ is the cluster persistence, and $P_0$ is the $\phi
\rightarrow 0^+$ limit of single bubbles. 
This equation would result in a  decay
exponent $\theta(\phi) = \theta(0) (1-\phi)$.  It is {\em a priori} 
unlikely that this drastic mean-field approximation,
or the resultant linearity, holds as an exact relationship. For example, 
in the droplet model \cite{Lee97} appreciable curvature corrections, 
expected to enter at $O[(1-\phi)^2]$, {\em must} develop since $\theta(0)=1$. 
Indeed, we see below that such non-mean-field effects occur in soap froths. 
However,  the {\it near\/} linearity of the froth data does imply that
correlation effects are small for cluster persistence in soap froths. 

To fit all of our froth data, we use an 
expansion in powers of $1-\phi$, making no assumptions about the
behavior near $\phi=1$, and obtain an excellent fit: 
$\theta(\phi) = -0.001^{\pm 0.003}+1.15^{\pm 0.02} (1-\phi) +0.15^{\pm 0.03} 
(1-\phi)^2$.  
 Higher powers in $(1-\phi)$ are not indicated, since we find
that the coefficients are zero within error bars.  
The quadratic term is  small but significantly non-zero --- 
this upward curving $\theta(\phi)$ indicates that the persistence decays,
 for all $0<\phi<1$, slightly slower than our 
mean-field argument would predict.  Evidently
cluster persistence reveals positive correlations between persistent
regions and neighboring cells of the same phase.
The variation near $\phi=1$ is linear within 
errors, as was conjectured in \cite{Lee97} by analogy with the droplet model.

We turn now to the question of the strong persistence decay in froths
relative to other curvature-driven systems, for which purpose we 
consider the standard persistence exponent, or equivalently the 
persistence of isolated bubbles.
The extrapolation to $\phi=0$ of our quadratic fit gives an improved
value of  
$\theta(0)=1.30 \pm 0.01$ for single bubbles \cite{previous},
which significantly disagrees with previous predictions 
of $\theta(0)=1$.
  In the scaling state, if a constant 
fraction of remaining bubbles have persistent cores, 
and if these cores have a non-decaying average area,
then the persistence is proportional to the density, and $\theta=1$.
Our result of $\theta\simeq 1.3$ implies either a continual
erosion of the persistent cores in bubbles, a continual decrease in the 
fraction of bubbles with persistent cores, or both.  By direct measurement
we find that the average persistent core size {\em does not decay at
late times,} while in contrast, the fraction of bubbles with persistent cores
steadily decays and entirely accounts for the exceptionally
strong persistence decay of individual bubbles in soap froths.
This strong decay is not shared by solutions of topological or droplet
mean field models (with or without vertices, respectively)
or by simulations of Potts models. 

\begin{figure}
\begin{center}
\hbox{\epsfxsize=3.0truein
\epsfbox{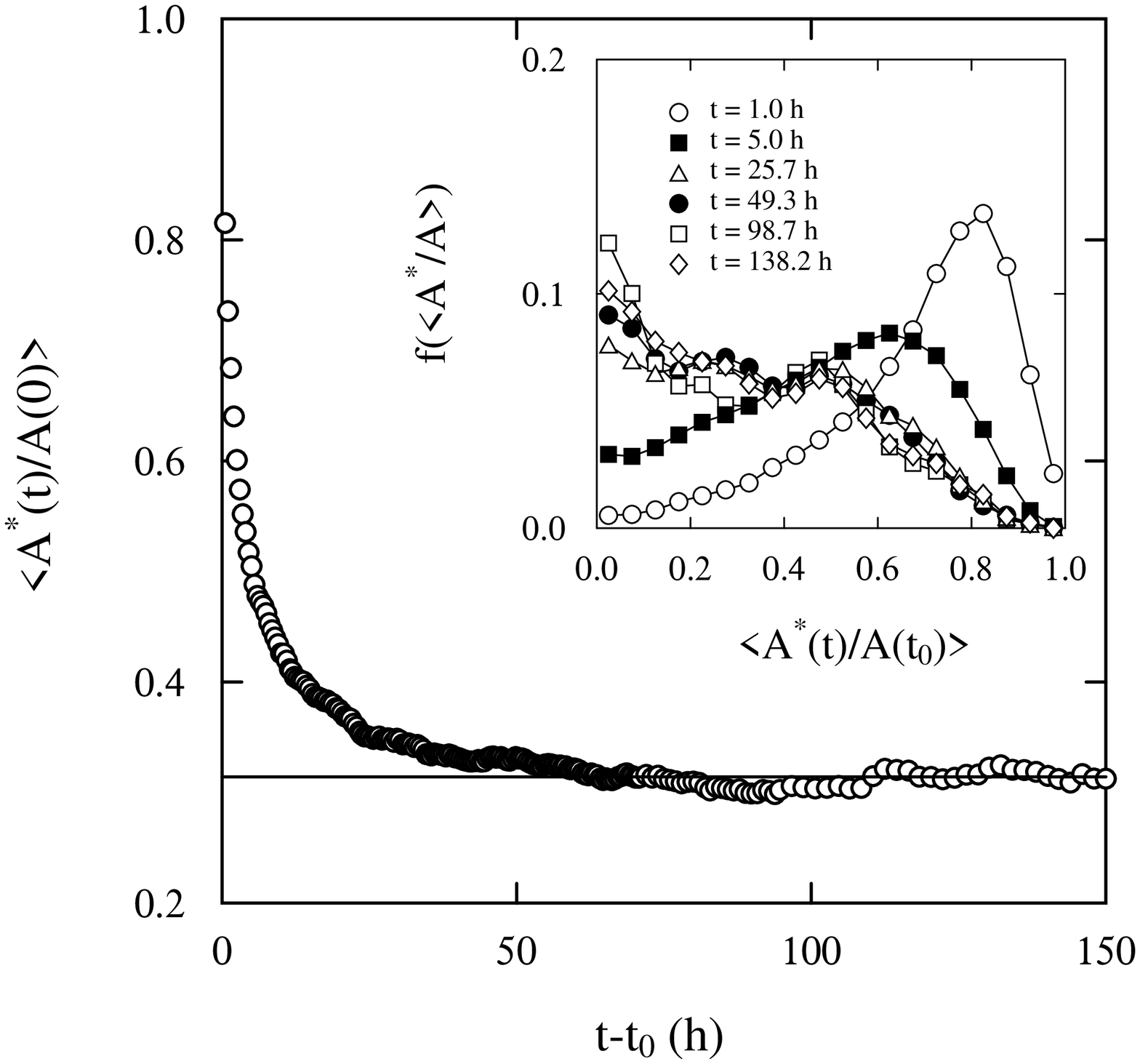}
\qquad
\epsfxsize=2.5truein
\epsfbox{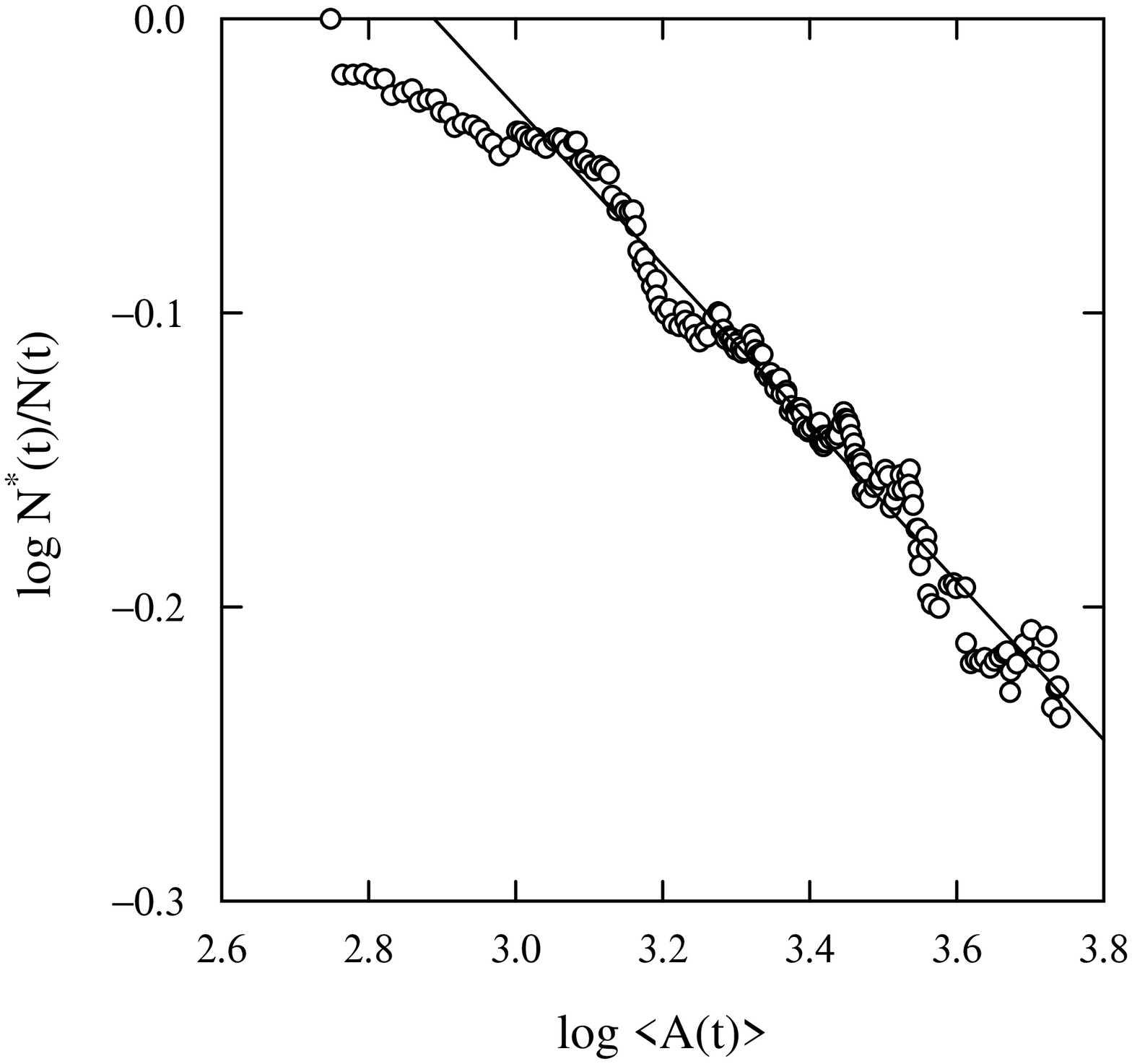}}
\end{center}
\caption{
Left: plot versus time of $\langle A^*(t)/A(t_0)\rangle$, the average of 
the persistent area within a bubble at time $t$ normalized by the entire 
area of the same bubble at $t_0$.
The averages are restricted to those bubbles with non-zero persistent regions.
The horizontal line, at $\langle A^*(t)/A(t_0)\rangle  = 0.314$ is the 
average value from $50-150$ h.  The distribution for various times is shown 
in the inset; it approaches a stationary form.
Right:  fraction of bubbles that contain persistent volume at time $t$
vs. the average bubble area. The best fit (solid line) gives a decay exponent
$-0.27 \pm 0.02$, where the error bar is approximate.}
\label{FIG:areapersist}
\end{figure}

We first measure the average persistent area $\langle A^*(t)/A(t_0) \rangle$,
where averages are restricted to the $N^*(t)$ bubbles that contain
persistent regions at time $t$, and we have normalized each bubble by its 
area at $t_0$.   
As shown on the left in fig.~\ref{FIG:areapersist},
it does not change in time after 50 h, and its distribution approaches a
time-independent form.  This is qualitatively similar to the evolution of
dilute droplet models \cite{Lee97}, where 
$\langle A^*(t) / A(t_0) \rangle \approx 1$ 
since all surviving drops have a  persistent core from $t_0$.

We next measure the number of bubbles that contain
any persistent area, $N^*(t)$, and plot the ratio $N^*/N$ vs
$\langle A(t) \rangle$ on the right in fig.~\ref{FIG:areapersist}.
While noisy, there is a clear decay, with a best-fit 
giving $N^* \sim \langle A \rangle ^{-1.27 \pm 0.02}$.
This is a dramatic and unexpected result.  
It implies a new length scale $L^* \sim t^{0.64 \pm 0.01}$, 
characterizing the separation of bubbles that have persistent volume, 
that grows faster than the bubble size $L \sim t^{1/2}$ \cite{Glazier92,Stavans93}.
It also independently 
implies a persistence decay $P_0^* = \langle A^*(t) \rangle N^* \sim N^*$,
indicated by the arrow in fig.~\ref{FIG:exppersist}, that is  
consistent with our previous result. 

\end{multicols}

\begin{multicols}{2}

The power-law decay of $N^*(t)/N(t)$ implies significant movement of the
center of mass of a bubble, since persistence within a given surviving
bubble cannot be entirely lost without the bubble moving off 
of its original center of mass.  Bubble motion on the scale 
of the bubble size is therefore directly implicated in the strong persistence 
decay of soap froths.

It is worthwhile to speculate on the causes of the bubble motion.
We have investigated the decay of persistence of individual bubbles, and
found numerous discrete decreases in the persistent fraction correlated with
significant and sudden motion of the bubble center of mass. This implicates
the correspondingly sudden
topological processes in bubble motion and hence in strong persistence decay.
$T2$ processes do not lead to sudden bubble motion for three-sided bubbles
as they vanish, and shrinking bubbles with four or five sides loose their
excess sides through $T1$ processes before they vanish. 
As a result, $T1$ processes are the main source of bubble motion.
Indeed, accurate numerical simulations of planar froths have seen that
bubble motion extends appreciable distances from $T1$ events \cite{Herdtle92},
in agreement with our observations.  It had been thought that $T1$ 
processes had no discernible effect on foam structure due to the unchanged
growth of bubble size when weak shear was applied \cite{Gopal95}. However
the growth of bubble size is well predicted by mean-field like arguments
\cite{Stavans93} that should become more accurate with shear due to mixing 
effects, and so is an incomplete probe of structure.  In contrast, our
results on persistence decay show that $T1$ processes, often not included
in theoretical work \cite{theory}, are important in the understanding of soap 
froth evolution. This is consistent with results of numerical modeling that show
that the width of the distribution of bubble sides, $\mu_2$, is dependent on 
the specific implementation of $T1$ processes \cite{Chae97}.

Potts models also have a cellular structure
and hence  topological processes occur throughout their evolution. 
There, within numerical accuracy 
$\theta'(0)=1$ \cite{Derrida96}, 
and so we must have both $\langle A^* \rangle$ 
and $N^*/N$ asymptotically constant and
hence no significant motion of bubbles. This
follows from the monotonicity of $\langle A^*(t) \rangle$ 
and $N \sim 1/\langle A \rangle \sim 1/t$.
What leads to this distinction between soap froths and $p \rightarrow \infty$
Potts models? Qualitatively, topological processes occur much more slowly in 
Potts models as compared with soap froths \cite{Glazier92}. However, the 
relative rate of $T1$ and $T2$ processes in Potts model evolution has 
neither been investigated nor compared with soap froths, where the ratio 
of $T1$::$T2$ is approximately $3::2$ \cite{Herdtle92}.  As yet, it is 
impossible to distinguish the role of the frequency of topological $T1$ 
processes from their speed in the strong decay of persistence.

In summary,
from $2d$ experiments of coarsening soap froths, we have measured persistence
decay exponents as a function of volume fraction. We find that soap
froths exhibit much faster persistence decay at all volume fractions
compared to other systems, including existing models of soap froth
evolution in which persistence has been measured.  Over time, a decreasing 
fraction of surviving bubbles contain any persistent area at all. This
implies significant motion of bubbles that is strongly
correlated with topological processes in the froth. 

While direct study of $T1$ processes is clearly desirable, 
we have demonstrated that the decay of persistence is a sensitive probe of the
{\em effect} of $T1$ processes. More 
fundamentally, persistence decay also allows the comparison of different
cellular systems, for instance between planar and bulk soap froths. 
In light of the possible proliferation of $T1$ processes in the late-time 
regime of bulk $3d$ froths \cite{Durian91}, it would be interesting to 
analyze persistence decay in those systems.  Such an exercise is now 
possible, in principle, 
with the development of spatially resolved MRI techniques \cite{Gonatas95}.

\centerline{***}

W.Y. Tam and K.Y. Szeto are supported by the Competitive Earmarked Research
Grant No.~HKUST6123/98P of the Research Grants Council of Hong Kong.
A. D. Rutenberg thanks the NSERC, and {\it le Fonds pour la Formation
de Chercheurs et l'Aide \`a la Recherche du Qu\'ebec} for financial support,
and J. Stavans for discussions and introductions.


\end{multicols}



\begin{thebibliography}{99}
\bibitem{endnote1} Our persistence approach is complementary to work on 
``survivors,''
or the statistics of the bubbles at an earlier time $t_0$
that will survive until a later time $t$
\cite{Levitan94,Tam98}. The topological distribution of survivors
is an exacting probe of soap froth models \cite{Levitan96}, and the area
contained in the survivors at $t_0$ is an upper bound on the persistent
area --- recovering $\theta \geq 1$ \cite{Levitan97}.
\bibitem{previous}
Previous work on the persistence decay of froths found
$\theta'(0)=1.16 \pm 0.02$ \protect\cite{Tam97} by fitting the average of
all of the single bubble decay curves. The resulting fit error 
is a significant underestimate. In this paper we extract error bars directly 
from the variances between different decay curves. For example, we directly
measure $\theta'(0)=1.38 \pm 0.11$, as shown in 
fig.~\protect\ref{FIG:exppersist}.
\end{thebibliography}
\end{document}